\documentclass[aip,numerical]{revtex4-1}

\usepackage{graphicx}
\usepackage{dcolumn}
\usepackage{bm}

\begin{document}

\preprint{}

\title[]{Impedance of the single electron transistor at radio-frequencies}

\author{C. Ciccarelli}
\email[cc538@cam.ac.uk]{}
\author{A.J.Ferguson}
\email[ajf1006@cam.ac.uk]{}

\affiliation{
Cavendish Laboratory, University of Cambridge, J. J. Thomson Avenue, Cambridge, CB3 0HE, United Kingdom.
}%

\date{\today}

\begin{abstract}
We experimentally characterise the impedance of a single electron transistor (SET) at an excitation frequency comparable to the electron tunnel rate. Differently from usual rf-SET operations, the excitation signal is applied to the gate of the device. At zero source-drain bias the single electron transistor displays both resistive (Sisyphus resistance) and reactive (tunnelling capacitance) components to its impedance. We study the bias dependence of the complex impedance, investigating its response as the electron tunnel rate becomes large with respect to the driving frequency. The experimental data are compared to values calculated from a master equation model.

\end{abstract}

\pacs{Valid PACS appear here}
\keywords{single electron transistor, rf-SET, Sisyphus resistance}
\maketitle
The single electron transistor (SET) is a fundamental nano-scale electronic device, it consists of an island coupled to source and drain leads by two low-capacitance tunnel junctions. \cite{ferry93} Its conductance is modulated by the polarization charge induced on the island by its electrostatic environment. As a result, it is possible to use the SET as an ultra-sensitive and high-bandwidth charge transducer, capable of detecting single electrons on a sub-microsecond timescale.\cite{schoelkopf:science98,korotkov:apl99,aassime:prl00,johansson:prl02,rimberg:nature03} An important characteristic that determines the performance of a SET is the rate with which it responds to an external perturbation. This is limited by the tunnelling rate of the electrons at the source and drain junctions and determines the ultimate charge detection bandwidth. When the time scale of the perturbation approaches the tunnelling time of the electrons, the SET response becomes susceptible to the stochastic nature of the tunnelling events. In this Letter we quantify the complex impedance that arises due to the competition between tunnelling rates and radio-frequency excitation.

The dissipative response (Sisyphus resistance) of a single electron box (SEB)\cite{persson:nl10} to a radio-frequency signal has previously been measured. A single electron device can also exhibit a capacitance due to electron tunnelling events leading, rather than being exactly in phase with, the excitation. This capacitance has been investigated in gallium arsenide quantum dots\cite{ashoori:prl92,cheong:apl02} and, similar to the Sisyphus resistance, allows measurements of single electron charging with just a single tunnel junction. This tunnelling capacitance is distinct from the quantum capacitance that arises from bandstructure curvature in single Cooper pair devices\cite{duty:prl05,roschier:prl05} and double quantum dots.\cite{petersson:nl10}

\begin{figure}[ht]
\centering
\includegraphics[angle=0,width=0.5\textwidth]{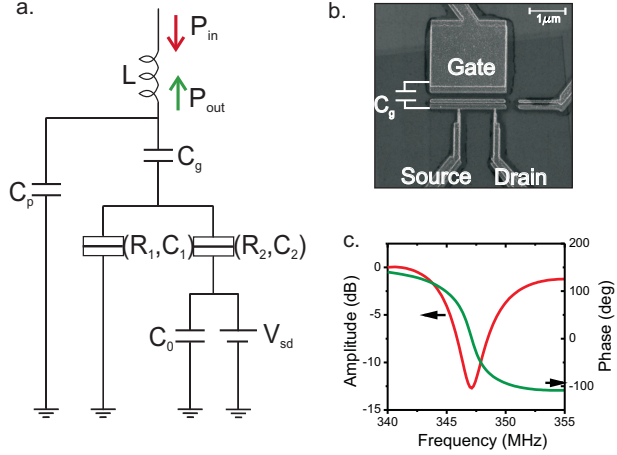}
\parbox{14cm}{\caption{(Color online) (a) A schematic of the resonant circuit. The SET is connected from the gate to the chip inductor ($L=560$ nH) which, together with a parasitic capacitance ($C_p=0.4$ pF), comprises the resonator. A bias-tee on the PCB allows us to apply a dc voltage to the gate. Chip capacitors ($C_{0}=150$ pF) are connected from the source and drain to ground to provide an rf ground. (b) Scanning electron micrograph of the aluminium SET. Sample resistance at low temperature is $2R=200 k\Omega$, where $R$ is the resistance of each tunneling junction. (c) Amplitude and phase characteristic of the reflected signal at $V_{sd}=0V$ as a function of the frequency.
}\label{figone}}
\end{figure}

Aluminium SETs were fabricated by a standard double angle evaporation technique using bilayer resist and controlled oxidation.\cite{fulton:prl87} They were measured in a dilution fridge at an electron temperature lower than 200~mK. A magnetic field (B=600~mT) was applied to suppress superconductivity in the aluminium. A radio frequency resonant circuit, consisting of a chip inductor, was connected to the \textit{gate} of the SET (Fig.~1(a) \& (b)). This is different from the usual configuration of the rf-SET\cite{schoelkopf:science98}, where the resonant circuit is connected to the source-drain of the SET, and where the largest contribution to the reflection coefficient is from modulation of the differential conductance.

The circuit was driven at resonance ($f_{0}=347$ MHz). The amplitude of the rf-signal sent to the device was $dn_{g}=C_{g}dV_{g}\sim 0.09~e$ (2xQx$\delta V_{g}$x$C_{g}$=2x60x2$\mu V$x0.1fF, where $\delta V_{g}$ is the amplitude of the rf-drive and Q is the resonator's Q-factor, estimated from the $3dB$ part of the resonance curve shown in Fig.\ref{figone}(c)). After amplification by a low temperature and room temperature amplifier, the reflected signal from the resonant circuit was mixed with the reference signal to provide phase sensitive detection. The output of the mixer was amplified, low pass filtered and digitised with an oscilloscope. By using a line stretcher we varied the phase of the incident signal to the device with respect to the reference signal. As a result, we obtain the in-phase and quadrature components of the reflected signal, allowing us to calculate the amplitude and phase response of the SET.

\begin{figure}[ht]
\centering
\includegraphics[angle=0,width=0.5\textwidth]{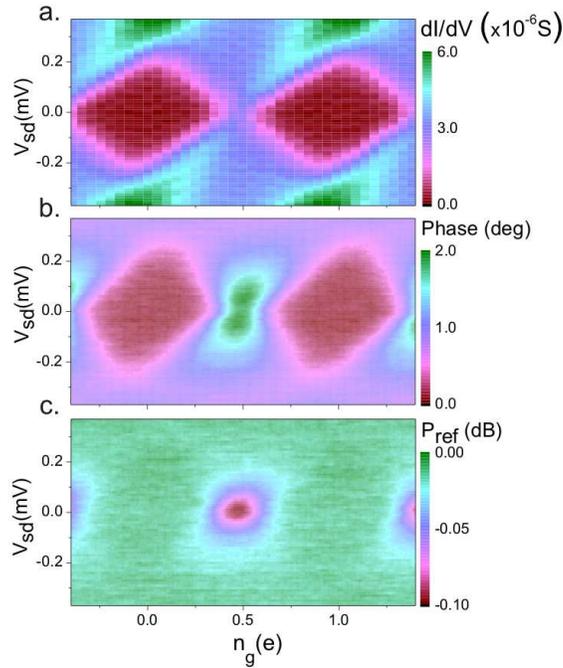}
\parbox{14cm}{\caption{(Color online) (a) The differential conductance of our SET is measured using a low-frequency lock-in amplifier. The total capacitance of the island $C_{\Sigma}=0.6$ fF, corresponds to a charging energy of $E_{c}=\frac{e^{2}}{2C_{\Sigma}}=0.13$ meV, the gate capacitance is $C_{g}=0.1$ fF. (b) Phase shift of the reflected signal. (c) Amplitude of the reflected signal relative to the refection in the blockaded region.}\label{figtwo}}
\end{figure}

We briefly compare the differential conductance of the SET, measured using a standard low-frequency lock-in technique, to the amplitude and phase response of the reflected signal (Fig.~\ref{figtwo}). In the phase response we can clearly observe the Coulomb diamonds that appear in the differential conductance graph, these corresponding to the blockade regime in which transport through the island is forbidden. Additionally there is a peak in the phase shift around the zero bias degeneracy points ($V_{sd}=0$~V, $n_g=(m+1/2)$~e). In the amplitude response, there is a dissipative signal at the degeneracy points that reduces with increasing source drain bias. Strikingly, by comparison with the phase response, the Coulomb diamonds are not observed.

\begin{figure}[ht]
\centering
\includegraphics[angle=0,width=0.5\textwidth]{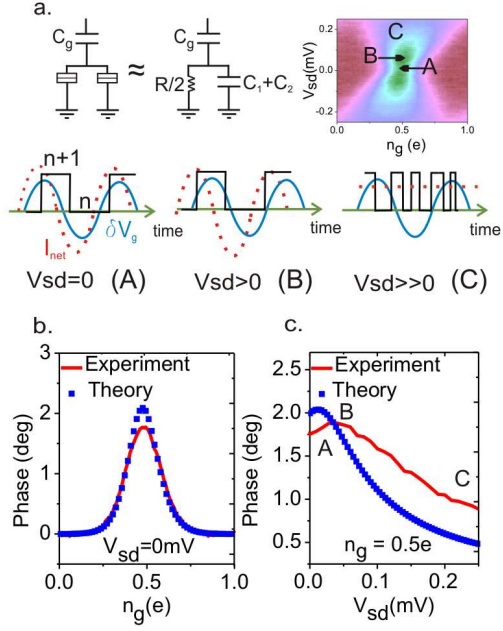}
\parbox{14cm}{\caption{(Color online) (a) The top left diagram shows the equivalent circuit for the two tunneling junctions in parallel. This includes the resistance and the geometrical capacitance of the junction. The top right image is a snapshot of the graph of the reflected signal phase as a function of the dc voltage applied to the gate and of the source to drain bias. On the graph three points are shown: A indicates the point of zero source to drain bias, B marks the peak in the phase shift at low source to drain biases and C indicates the region of high source to drain biases. The three diagrams at the bottom show the tunnelling events on the time scale of the rf-drive for each of these three points. $I_{net}$ is the net current to ground that results from the tunnelling of the electrons. We see that, as $V_{sd}$ is increased (B), the tunnelling events happen "early" in the rf-cycle and $I_{net}$ develops a negative phase shift with respect to the drive. If $V_{sd}$ is further increased (C), however, $I_{net}$ is generated by stochastic tunnelling events no longer correlated to the drive.  (b) and (c) Graph of the measured and calculated variation of the reflected signal phase shift as a function of the dc gate voltage (b) and of the source to drain voltage (c).}\label{figthree}}
\end{figure}

We now analyse the phase response, starting with observation of the Coulomb diamonds. When the device is in Coulomb blockade (CB), the gate capacitance ($C_{g}$) and the junction capacitances ($C_{1}+C_{2}$) are in series (Fig.\ref{figthree}(a)), with little contribution to the impedance from the tunnel junction resistance. Outside the blockaded region, the capacitance becomes shunted by the junction resistance. This results in an increase in capacitance ($\frac{C_{g}(C_{1}+C_{2})}{C_{g}+C_{1}+C_{2}}$) seen by the resonator, lowering the resonant frequency, and explaining the phase shift.

Near the zero bias degeneracy points a description of the electron tunnelling dynamics is necessary. With the rf signal superimposed on the gate, the island cyclically transits between the nearly degenerate $n$ and $n+1$ charge states. If the tunnel rates at zero instantaneous rf bias (we estimate at the zero bias degeneracies $\Gamma=\frac{kT}{e^2R}\sim$2~GHz, where $R$ is the resistance of one tunnelling junction), are greater than the drive frequency, the electrons on average tunnel early in the rf cycle. This phase shifts the current with respect to the rf voltage, leading to an effective capacitance $C_{eff}$ in parallel with the SET. The total capacitance $C$ can then be written as $C=\frac{(C_{1}+C_{2})C_{g}}{C_{\Sigma}}+\frac{C_{g}}{C_{\Sigma}}\frac{d<en>}{dV_{g}}$, where $<en>$ is the average charge on the island. This expression can be found from the expression of the polarisation charge induced on the island by the gate $Q_{g}=en-[C_{1}(V_{sd}-V_{i})-C_{2}V_{i}]$, where $V_{i}$ is the island potential. The first term represents the dc limit of the capacitance, obtained by considering $C_{g}$ in series with the capacitances of the source and drain tunnelling junctions. The second term represents the contribution from $C_{eff}$ and can be written as $-\frac{C_{g}}{C_{\Sigma}}\frac{e\dot{P_{n}}}{(1/C_{g})\frac{dn_{g}}{dt}}$, where $(1/C_{g})dn_{g}$ is the amplitude of the rf drive and $P_{n}$ is the probability of having $n$ electrons on the island. When $k_{B}T<<E_{c}$ and $dn_{g}<<e$, tunnelling to higher energy states can be neglected and $P_{n}$ is found by solving the Master equation which involves the states $n$ and $n+1$\cite{ferry93}

\begin{equation}\dot{P_{n}}=\Gamma_{n+1,n}P_{n+1}-\Gamma_{n,n+1}P_{n}
 \end{equation}
 $$\dot{P_{n+1}}=\Gamma_{n,n+1}P_{n}-\Gamma_{n+1,n}P_{n+1}$$

$\Gamma_{n+1,n}$ and $\Gamma_{n,n+1}$ are the tunnelling rates of the island between the two states $n$ and $n+1$\cite{grabert}. An increase in $V_{sd}$ leads to an increase in the tunnelling rate of the electrons at the drain lead. We therefore expect the tunnelling events to happen "earlier" with respect to the case of zero bias and the current to be more phase shifted with respect to the rf signal (Fig. \ref{figthree}(a)). Accordingly, a maximum in phase (maximum $C_{eff}=0.03$~fF) occurs at $V_{sd}=30~\mu$V (Fig.\ref{figthree}(c)). When $V_{sd}$ is increased above this value the contribution of the source to drain potential becomes dominant, the tunnelling events are no longer sensitive to the rf drive, therefore leading to a decrease in the phase shift. In Fig. \ref{figthree}(b) and (c) we report the calculated phase shift as a function of both $n_{g}$ and $V_{sd}$. The simulations have been performed at a finite temperature of $120mK$ and with $dn_{g}=0.09e$. By comparing the simulations with the experimental results, we observe that there is a good agreement in the $n_{g}$ dependence. The $V_{sd}$ dependence qualitatively agrees with the experimental observations, but presents some discrepancies that could be fitted by using a higher value for the temperature or $dn_{g}$. We could not explain, however, the origin of this disagreement.
\begin{figure}[ht]
\centering
\includegraphics[angle=0,width=0.5\textwidth]{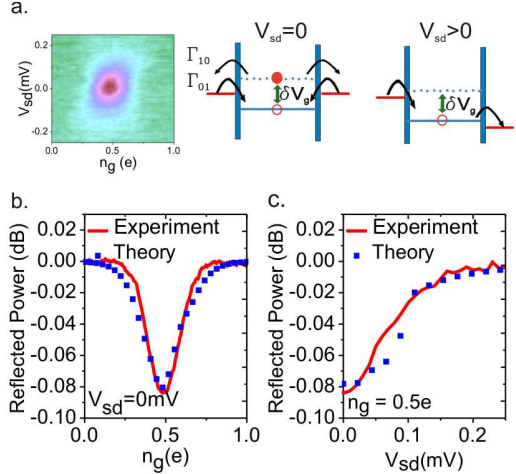}
\parbox{14cm}{\caption{(Color online) (a) Reflected signal amplitude as a function of the $V_{sd}$ and $n_{g}$. The decrease at $V_{sd}=0$ V and $n_{g}=0.5$~e is explained with an increase in energy dissipation by the SET due to the Sisyphus tunnelling processes. The top right diagrams show the different tunnelling processes involved at zero and positive $V_{sd}$. In the first case, the Sisyphus tunnelling events lead to the absorption of extra energy from the resonator and thus to a decrease in the reflected  power amplitude. In the second case, the energy needed for the tunnelling is provided by the battery and no decrease in amplitude is observed. (b) \& (c) Graph of the measured and calculated variation of the reflected power amplitude as a function of the dc gate voltage (b) and of the source to drain voltage (c). }\label{figfour}}
\end{figure}

We next consider the behavior of the amplitude. From Fig.~\ref{figfour}(b) and (c), we observe that the reflected (dissipated) signal reaches a minimum (maximum) at $n_{g}=0.5$~e and $V_{sd}=0$~V. When $V_{sd}=0$~V, the SET effectively behaves as a single electron box (SEB) and its dissipative response to a radio-frequency signal can be understood in terms of the Sisyphus resistance.\cite{persson:nl10} On average electrons tunnel after the degeneracy point between the lead and island chemical potentials is passed, therefore energy dissipation occurs in each half of the rf-cycle, dissipating energy from the resonator. From the average power $\overline{P}$ dissipated by the SET (Fig.~\ref{figfour}), we deduce the value of the effective resistance $R_{eff}$, $R_{eff}=((1/C_{g})dn_{g})^{2}/2\overline{P}$. Our resonator is lossy ($Q\sim 60$), so we find $\overline{P}$ by measuring the relative variation of the reflected power. At $V_{sd}=0$~V, the effective resistance reaches a minimum value $R_{eff}=16 M\Omega$.

As $V_{sd}$ becomes greater than the rf amplitude, the tunnelling events onto and off the island are driven largely by the dc bias rather than the rf excitation (Fig.~\ref{figfour}(a))and the contribution from the Sisyphus dissipation becomes smaller. The absence of the CB diamonds in Fig.~\ref{figtwo}(c) shows that the amplitude response is completely dominated by the Sisyphus dissipation, which overshadows the effects of the SET resistance variation at higher values of $V_{sd}$.
The average power dissipated in one period of the rf drive can be calculated as

\begin{equation}
\overline{P}=\frac{\omega}{2\pi}\int_{0}^{\frac{2\pi}{\omega}}(E_{n+1}-E_{n})\dot{P_{n}}\theta[(E_{n+1}-E_{n})\dot{P_{n}}]dt
\end{equation}

where $E_{n+1}-E_{n}$ is the energy difference between $n+1$ and $n$ states. The Heaviside step function ($\theta$) is ensures that the transitions between the states $n$ and $n+1$ are energetically allowed. There is close agreement between the theory and the experimentally measured amplitude response as a function of $n_{g}$ and $V_{sd}$ (Fig.~\ref{figfour}(b) \& (c)).

\begin{figure}[ht]
\centering
\includegraphics[angle=0,width=0.5\textwidth]{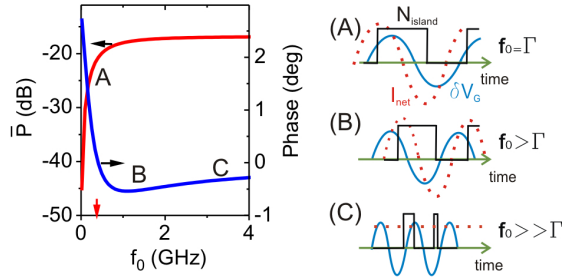}
\parbox{14cm}{\caption{(Color online) Calculated variation of the phase shift and of the power dissipated in the device, $\overline{P}$, as a function of the ac excitation frequency $f_{0}$ for $n_{g}=0.5$e and $V_{sd}=0$V. The arrow indicates the frequency at which we worked in our experiment. On the graph of the phase shift three points are shown: A indicates the region in which the electrons' tunnelling rate is comparable to the frequency of the rf-drive, B marks the minimum value of the phase shift, which is reached at higher excitation frequencies, and C indicates the region in which the frequency of the excitation is much higher than the electrons' tunnelling rate. The three diagrams on the right show the tunnelling events involved in each of these three cases on the time scale of the rf-drive.}\label{figfive}}
\end{figure}

We expect that the reactive and dissipative components of the SET response to an rf-drive will strongly depend on the frequency of the excitation. In our experiment we don't have the flexibility to make a frequency dependent study of this response, however, we show in Fig.~\ref{figfive} the simulated variation of the reflected signal phase shift and of the dissipated power, $\overline{P}$, as a function of the excitation frequency. As the frequency of the drive becomes much higher than the electrons' tunnelling rate, the tunnelling events are random and no longer correlated to it. However, the probability for them to occur is higher at the maximum amplitude of the rf-drive as the tunnelling rate is higher. Accordingly, the power dissipated in the device saturates towards a maximum value at higher frequencies. $C_{eff}$, on the other hand, tends to zero and the phase shift of the reflected signal is only determined by $\frac{(C_{1}+C_{2})C_{g}}{C_{\Sigma}}$. It is interesting to note that for frequencies higher than our working frequency (347 MHz), a negative phase shift is expected, opposite to what we observe. In diagram B. of Fig.~\ref{figfive}, this is explained with the tunnelling events happening in the second half of the rf-cycle, resulting in a positive shift of the net current with respect to the drive.

By connecting the gate, rather than the source or drain, to the radio-frequency circuit we are able to determine the Sisyphus resistance and tunnelling capacitance contributions to the SET impedance. These effects will also be present in a conventionally measured rf-SET (resonator on source-drain) and should be included if the impedance needs to be accurately known.

We are grateful for helpful discussions with Andrew Armour and financial support under the EU Grant FP7-214499 NAMASTE and from Hitachi Cambridge Laboratory. A.J.F. acknowledges the support of a Hitachi research fellowship.

\end{document}